\documentstyle[aps,prl,twocolumn,amsmath,epsf]{revtex}

\begin{document}
\draft
\wideabs{
\title{Three-body problem in Fermi gases with short-range
interparticle interaction}
\author{D.S. Petrov}
\address{FOM Institute for Atomic and Molecular
Physics, Kruislaan 407, 
1098 SJ Amsterdam, The Netherlands \\
Russian Research Center, Kurchatov Institute, 
Kurchatov Square, 123182 Moscow, Russia}

\date{\today}
\maketitle
\begin{abstract}
We discuss 3-body processes in ultracold two-component Fermi
gases with short-range intercomponent interaction
characterized by a large and positive scattering length $a$.
It is found that in most cases the probability of 3-body
recombination is a universal function of the mass ratio and
$a$, and is independent of short-range physics. We also
calculate the scattering length corresponding to the
atom-dimer interaction. 
\end{abstract}
}
\narrowtext

In the search for fermionic superfluidity, ultracold Fermi
gases with strong interactions and high densities are
produced routinely. In current experiments with $^{40}$K
\cite{Jin} and $^6$Li \cite{Ketterle,Grimm,Thomas,Salomon},
the
scattering length $a$, corresponding to the interaction
between different fermionic species, has been modified using
a
powerful tool of Feshbach resonances.
Reaching high densities is limited by 3-body recombination
-- the process in which two atoms form a
bound state and a third one carries away the binding energy
$\varepsilon$. In the case of a two-component Fermi gas the
3-body process requires at least two identical fermions to
approach each other to a distance of the order of
the size of the final bound state. Therefore, in the
ultra-cold limit the recombination probability acquires an
additional small factor
$K/\varepsilon$, where $K$ is the relative kinetic energy of
identical fermions (cf. \cite{Esry}). One may speculate
that limitations on achieving high densities are not as
severe as in Bose gases, where the recombination rate is
independent of the kinetic energy of particles.

Theoretical studies of the 3-body problem have
revealed the existence of two universality classes for the
case where $a$ greatly exceeds the characteristic radius of
interatomic interaction, $R_e$ \cite{EfimovMR,EfimovSR}. For
the first class,
short-range physics is not important and the 3-body problem
can be described in terms of 2-body scattering lengths and
masses of particles. One can then use the zero-range
approximation for the interatomic potential, which has been
successfully demonstrated, e.g. in the calculations of
neutron-deuteron scattering with the total spin $S=3/2$
\cite{Ter}. For the second class, where short-range
behavior is important, the description of the 3-body
problem requires at least one parameter coming from
short-range physics \cite{Danilov}. For two identical
fermions ($\uparrow$) interacting with a third particle
($\downarrow$), the presence of two universality classes
has been shown by Efimov \cite{EfimovMR,EfimovSR}. For a
large $a$, the pair interaction between distinguishable
particles leads to the appearance of an attractive $1/r^2$
interaction in the 3-body system \cite{note}. If the mass
ratio $m_\uparrow/m_\downarrow$ is smaller than
approximately 13, then this attraction is not sufficient to
overcome the centrifugal barrier
$l(l+1)/r^2$ ($l=1$ due to the symmetry). In this case the
probability of all three particles being in the volume
$R_e^3$ vanishes and short-range physics drops out of the
problem. However, the effective attraction increases with
the mass ratio and for $m_\uparrow/m_\downarrow\agt 13$ we
have a well-known phenomenon of the fall of a particle into
the center in an attractive $r^{-2}$ potential \cite{LLQ}.
The shape of the wavefunction at distances of the order of
$R_e$ then significantly influences the large-scale
behavior and short-range parameters of the interaction
potential are required to describe the 3-body system.  

In this paper we discuss a 3-body system
($\uparrow\uparrow\downarrow$) containing two identical
fermions and belonging to the first universality class. In
the ultra-cold limit we find universal functions of $a$ and
$m_\uparrow/m_\downarrow$ for the probability of 3-body
recombination and for the amplitude of atom-dimer
scattering. As expected, the recombination to comparatively
deep bound states is much slower than in the Bose case.
However, despite the suppression factor of $K/\varepsilon$,
the recombination to a weakly bound $s$-level ($a>0$ and
$a\gg R_e$) for realistic parameters of a two-component
Fermi gas can be as important as in a Bose gas with the
same density and scattering length.

In the center of mass reference frame the state of the
3-body system ($\uparrow$ $\uparrow$
$\downarrow$) with total energy $E$ is described by the
Schr\"odinger equation
\begin{equation}\label{Schrrescaled}
\left[-\nabla_X^2-E\right]\Psi=-\sum_\pm
V\left(\frac{\hbar({\bf
x}\tan\theta\pm {\bf
y})}{2\sqrt{m_\uparrow}}\right)\Psi,
\end{equation}
where $\hbar{\bf y}/\sqrt{m_\uparrow}$ is the distance
between identical
$\uparrow$-fermions, $\hbar{\bf x}/\sqrt{2\mu}$ is the
distance between their center of mass and the
$\downarrow$-particle,
$\mu=2m_\uparrow m_\downarrow/(2m_\uparrow+m_\downarrow)$
is the corresponding reduced mass, and
$\theta=\arctan\sqrt{1+2m_\uparrow/m_\downarrow}$. The
vector $X=\{{\bf x},{\bf y}\}$ describes the rescaled 6D
configuration space for the 3-body problem. The potential
of interaction between distinguishable particles is $V$ and
the interaction between identical fermions is omitted.

We assume that the range $R_e$ of the interatomic potential
$V$ is much smaller than the (positive) scattering length
$a$, which in the rescaled coordinates equals
$1/\sqrt{\varepsilon}$. Therefore, in the 6D space the rhs
of Eq.(\ref{Schrrescaled}) vanishes everywhere except in
the close vicinity to 3D subspaces $S_+$ and $S_-$, which
we define parametrically as $X_\pm({\bf r})=\{{\bf
r}\cos\theta,\pm{\bf
r}\sin\theta\}$, where the 3D vector ${\bf r}$ is a
parameter. Near the subspace $S_+$ ($S_-$) the
$\downarrow$-particle is close to one of the
$\uparrow$-fermions and is separated by the distance ${\bf
r}$ from the other. The distance to the first
$\uparrow$-fermion we denote by ${\bf r_\bot}$, which is
then a parameter in a 3D subspace orthogonal to $S_+$
($S_-$). The subspace $S_-$ differs from $S_+$ by the
permutation of the
$\uparrow$-fermions (${\bf y}\rightarrow -{\bf y}$). We thus
face a typical boundary-value problem for the Poisson
equation
$\left[-\nabla_X^2-E\right]\Psi=0$, with a boundary
condition given by
\begin{equation}\label{Boundary}
\Psi\sim 1-1/(\sqrt{\varepsilon}|{\bf r_\bot}|)+O(\bf
r_\bot).
\end{equation}

Introducing functions $f_+({\bf r})$ and
$f_-({\bf r})$ we can write a general solution of
Eq.(\ref{Schrrescaled}), which is valid everywhere except
the narrow vicinity of $S_+$ and $S_-$:
\begin{equation}\label{Psi}
\Psi=\Psi_0\left(X\right)+\sum_\pm \int G_E\left(\left|
X-X_\pm({\bf r})\right|\right)f_\pm({\bf r})d^3{\bf r}.
\end{equation}
Here $\Psi_0$ is a solution of the Poisson equation and has
no singularities at $S_+$ or $S_-$. We restrict $\Psi_0$ to
be finite everywhere and, hence, for $E<0$ we have
$\Psi_0\equiv 0$. The
Green's function $G_E$ is a solution of
Eq.(\ref{Schrrescaled})
with the rhs $\delta^6(X)=\delta^3({\bf
x})\delta^3({\bf y})$ and is given by
\renewcommand{\arraystretch}{1.5} 
\begin{equation}
G_E(X)=\left\{\begin{array}{lr}\frac{-EK_2(\sqrt{-E}\left|X\right|)}{8\pi^2\!
X^2},&E<0\\
\frac{iEH_2(\sqrt{E}\left|X\right|)}{16\pi^3
X^2},&E>0\end{array}\right.
\xrightarrow[E\rightarrow 0]{}
\frac{1}{4\pi^3 X^4}.\nonumber 
\end{equation}
\renewcommand{\arraystretch}{1}

\noindent Here $K_2$ is an exponentially decaying
Bessel function and $H_2$ is a Hankel function representing
the outgoing wave.

In order to satisfy the symmetry with respect to permutation
of identical fermions we set $f_+({\bf
r})=-f_-({\bf r})=f({\bf r})$
and require $\Psi_0({\bf x},{\bf y})=-\Psi_0({\bf x},-{\bf
y})$. Eqs. (\ref{Boundary}) and (\ref{Psi}) then give
the integral equation for the function $f({\bf r})$
\begin{equation}\label{main}
\left(\hat{L}_E-\sqrt{\varepsilon}+\sqrt{-E}\right)f({\bf
r})=4\pi\Psi_0(X_+({\bf r})),
\end{equation}
where the hermitian operator $\hat{L}_E$ is given by
\begin{eqnarray}\label{LE}
\hat{L}_Ef({\bf r}) & = & 4\pi\int \Bigl[ G_E\left(|{\bf
r}-{\bf
r'}|\right)[f({\bf r})-f({\bf r'})]\nonumber \\
 & + & G_E\left(\sqrt{{\bf r}^2+{\bf r'}^2-2{\bf r}{\bf
r'}\cos 2\theta}\right)f({\bf
r'})\Bigr]d^3{\bf r'}.
\end{eqnarray}
This operator conserves angular momentum, and we
can expand the solution of Eq.(\ref{main}) in
spherical harmonics and deal only with a set of uncoupled
integral equations for functions of a single variable $r$.
Eq.({\ref{main}) is a particular case of the Faddeev
equations
\cite{Faddeev} and in the momentum representation it was
first obtained in \cite{Ter} for $m_\uparrow=m_\downarrow$.

The method presented above allows us to solve the problem of
atom-molecule scattering for an arbitrary mass ratio. Let us
assume that the relative kinetic energy of a
$\uparrow\downarrow$-dimer and $\uparrow$-fermion is much
less than the dimer binding
energy $\varepsilon$. The total energy of the system is
then negative $E\approx -\varepsilon$ and, consequently,
$\Psi_0\equiv 0$. As the size of the dimer is much smaller
than
the inverse relative momentum, the $s$-wave contribution
dominates and Eq.({\ref{main}) becomes
\begin{equation}\label{Atom-Mol}
\hat{L}_{-\varepsilon}^{l=0}f(r)=0.
\end{equation}
Here $\hat{L}_{-\varepsilon}^{l=0}$ is an integral
operator which is obtained from $\hat{L}_{-\varepsilon}$
by integrating over angles.
 
From Eq.(\ref{Psi}) we find that far from the origin ($r\gg
1/\sqrt{\varepsilon}$) and in the region of $S_\pm$ the
wavefunction $\Psi\approx
\pm f(r)\exp(-\sqrt{\varepsilon}r_\bot)/4\pi r_\bot$.
Therefore, at these distances the function $f(r)$ describes
the atom-molecule relative motion and behaves
as $1-\beta/r$. The atom-molecule scattering length
is then given by $a_m=a\beta\sqrt{\varepsilon}\sin 2\theta$.
The
ratio $a_m/a$ is plotted in Fig.\ref{AA} as a function of
$m_\uparrow/m_\downarrow$. In the limit of
$m_\uparrow/m_\downarrow\gg1$ one can use the
Born-Oppenheimer
approximation. In
this case the heavy $\uparrow$-fermions move slowly in a
field produced by the exchange of the fast light
$\downarrow$-particle. The adiabatic behavior assumes
decomposition of the wavefunction into two parts. The first
part describes $s$-wave scattering of the heavy fermions.
The
motion of the light particle is, therefore,
described by a wavefunction antisymmetric with respect to
their permutation. At large distances ($y\gg
1/\sqrt{\varepsilon}$) the effective interaction has the
form
of a repulsive Yukawa potential $U(y)=\cos\theta
(m_\uparrow/m_\downarrow)
2\sqrt{\varepsilon}\exp(-2\cos\theta\,
y\sqrt{\varepsilon})/y$. The corresponding scattering
length is plotted in Fig.\ref{AA} as a dashed line. We
estimate
$a_m/a\sim\ln(m_\uparrow/m_\downarrow)$. 
\begin{figure}
\hspace{-0.2cm}
\epsfxsize=\hsize
\epsfbox{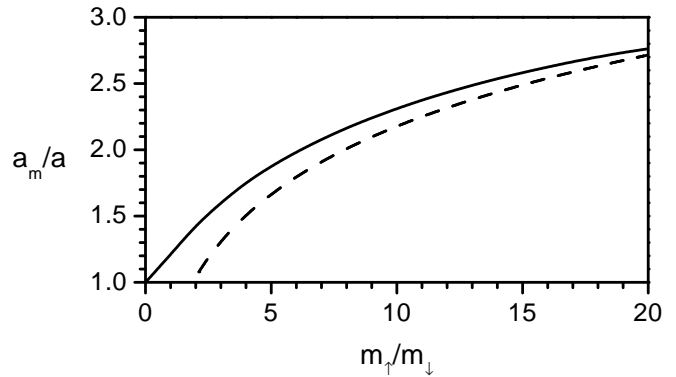}
\caption{\protect
The ratio $a_m/a$ versus $m_\uparrow/m_\downarrow$
calculated from Eq.(\ref{Atom-Mol}) (solid line) and from
the Born-Oppenheimer approximation (dashed
line).}
\label{AA}
\end{figure}

It can be shown analytically that $f(r)\sim r^\gamma$ near
the origin. The exponent $\gamma$ increases
monotonically from $1$ to $3$, with
$m_\uparrow/m_\downarrow$
growing from $0$ to $\infty$ indicating an effective
repulsion in the system \cite{EfimovMR}. In the case
of $m_\uparrow=m_\downarrow$ we obtain $a_m/a\approx 1.2$
(cf.
\cite{Ter}).

We now turn to the problem of 3-body recombination just
above the threshold $0<E\ll\varepsilon$. This process
($\uparrow+\uparrow+\downarrow\,\longrightarrow\,\uparrow+\uparrow\downarrow$)
takes place at interparticle distances
comparable with the size of a $\uparrow\downarrow$-dimer
($\sim 1/\sqrt{\varepsilon}$). Therefore at these distances
the shapes of the functions $\Psi$ and $f$ are independent
of $E$.
The normalization coefficient, however, depends on the large
scale behavior and, therefore, on energy $E$. We can
illustrate this by writing the wavefunction for free motion
in the form $\Psi_0({\bf x},{\bf
y})=\sin({\bf
k}\cdot{\bf y})\exp(i{\bf q}\cdot{\bf x})$, where the
momenta ${\bf k}$ and ${\bf q}$ satisfy the equation
$E=k^2+q^2$. At
distances comparable with $1/\sqrt{\varepsilon}$ the
function $\Psi_0$ reduces to ${\bf k}\cdot{\bf y}$, giving
rise
to the $k^2$ threshold law for the probability of 3-body
recombination \cite{Esry}. Considering the limit
$E\rightarrow 0$ we keep the term $4\pi{\bf
k}\cdot{\bf
r}\sin\theta$ in the rhs of Eq.(\ref{main}).

We now search for the solution of Eq.(\ref{main}) in the
form
$f({\bf r})=f(r){\bf k}\cdot{\bf r}/kr$, which corresponds
to the angular momentum $l=1$. Eq.(\ref{main}) then
transforms to
\begin{equation}\label{mainfinal}
\left(\hat{L}_0^{l=1}-\sqrt{\varepsilon}\right)f(r)=4\pi
kr\sin\theta.
\end{equation} 
An explicit form of the operator $\hat{L}_0^{l=1}$ can be
obtained from Eq.(\ref{LE}) by integrating out the angular
part
of $f({\bf r})$. 

The operator $\hat{L}_0^{l=1}$ has remarkable properties
which allow us to solve Eq.(\ref{mainfinal}) analytically.
{\it
Firstly}, for $f(r)=r^\nu$, where $\nu$ is in the region
$-4<{\rm
Re}\,(\nu)<2$, the action of the operator reduces to
%\begin{equation}\label{Lpower}
$\hat{L}_0^{l=1}r^\nu=\lambda(\nu)r^{\nu-1}$,
%\end{equation} 
and the function $\lambda(\nu)$ is given by
\begin{eqnarray}%\label{lambda}
&&\lambda(\nu)=\frac{\nu(\nu+2)}{\nu+1}\cot\frac{\pi\nu}{2}\nonumber\\
&&-\frac{\nu\cos2\theta\cos\left[(\nu+1)(2\theta-\pi/2)\right]+\sin\left[\nu(2\theta-\pi/2)\right]}{(\nu+1)\cos^22\theta\sin2\theta\sin(\pi\nu/2)}\nonumber.
\end{eqnarray}
In the specified interval of $\nu$ the
function $\lambda$ has two roots:
$\nu_0$ and $-\nu_0-2$. With $m_\uparrow/m_\downarrow$
increasing from 0 to approximately $13.6$
the root $\nu_0$ decreases from 1 to -1 approaching
the second root. At larger mass ratios the roots are
complex conjugates with the real part equal to -1.
Zeros of $\lambda(\nu)$ are important to
understand the behavior of $f(r)$ at short distances.
%Outside the strip $-4<{\rm
%Re}\nu<2$ the integral in $\hat{L}_0'r^\nu$ does not
%converge. 

{\it Secondly}, any eigenfunction $\chi(r)$ of the operator
$\hat{L}_0^{l=1}$ corresponding to the eigenvalue 1 can be
rescaled to generate an eigenfunction corresponding to a
new eigenvalue $p$ (for any $p>0$)
\begin{equation}\label{Scaling}
\hat{L}_0^{l=1}\chi(pr)=p\chi(pr).
\end{equation}

As $\lambda$ has two roots, the spectrum
of $\hat{L}_0^{l=1}$ is two-fold degenerate, and
for $p=1$ there is only two linearly independent
eigenfunctions which behave as $\chi_1\sim r^{\nu_0}$ and
$\chi_2\sim
r^{-\nu_0-2}$ at small $r$.
The complex exponents $\nu_0$ and $\nu_0^*$ at
larger mass ratios imply that the function
$\chi_1=\chi_2^*$ rapidly oscillates as $r\rightarrow 0$
and the Efimov effect takes place. 
In both cases we choose two real eigenfunctions which
at large distances
($r\gg 1$) have the
asymptotes $\chi_{1,2}(r)\approx 2\sin(r+\delta_{1,2})/r$.
The phases $\delta_1$ and
$\delta_2$ can be determined from
a numerical calculation of
Eq.(\ref{Scaling}).

The case $p=\sqrt{\varepsilon}$ corresponds to the collision
of a
molecule with an atom just below the 3-body recombination
threshold ($E=-0$) where the function $\Psi_0$ is zero.
Strictly speaking this process is described by the linear
combination
$f_{mol}(r)=A\chi_1(\sqrt{\varepsilon}r)+B\chi_2(\sqrt{\varepsilon}r)$.
The determination of the coefficients $A$ and $B$ involves
short-range physics, i.e. the knowledge of the 3-body
wavefunction $\Psi$ at distances of
the order of $R_e$, and is beyond the scope of this paper.
However, the matching procedure implies that at
these distances both terms in $f_{mol}(r)$ are of the same
order of magnitude, so $B/A\sim
(\sqrt{\varepsilon}R_e)^{2\nu_0+2}$. In the case of
$m_\uparrow/m_\downarrow<13.6$, the exponent $2\nu_0+2$ is
real and positive, and at distances $r\gg R_e$ one has
$f_{mol}(r)\approx A\chi_1(\sqrt{\varepsilon}r)$.
Hence, short-range
3-body parameters are unnecessary. 

Fortunately, the discussed case of
$m_\uparrow/m_\downarrow<13.6$ covers almost all practical
situations. Using
the function $\chi_1(r)$
and the scaling property (\ref{Scaling}) we can construct a
complete orthonormal set of functions $f_p(r)=p\chi_1(pr)$.
Completeness follows from the
equality $rf_p(r)=pf_r(p)$ and from the normalization
condition $\int_0^\infty
f_p(r)f_{p'}(r)r^2dr=2\pi\delta(p-p')$. The operator
$\hat{L}_0^{l=1}-\sqrt{\varepsilon}$ in Eq.(\ref{mainfinal})
can now be inverted and after some manipulation we obtain
\begin{eqnarray}\label{answer}
f(r)&=&-\frac{4\pi
k\sin\theta}{\varepsilon}\Biggl(\sqrt{\varepsilon}r+\lambda(1)+\frac{\lambda(0)\lambda(1)}{\sqrt{\varepsilon}r}\nonumber\\
&-&\frac{\lambda(0)\lambda(1)}{\sqrt{\varepsilon}r}\sqrt{\frac{\lambda(-1)}{2\pi}}\int_0^\infty\frac{\chi_1(z)zdz}{z-\sqrt{\varepsilon}r}\Biggr).
\end{eqnarray}
The integral in Eq.(\ref{answer}) is taken as a principal
value. The
solution $f$ is not singular at short distances. The terms
containing $r^{-1}$, $r^0$ and $r$ vanish as expected
except in situations where $\lambda(1)$, $\lambda(0)$ or
$\lambda(-1)$ are equal to 0. At distances $r\ll
1/\sqrt{\varepsilon}$ we have $f(r)\sim r^2$ and the
short-range physics does not come into play. For large
distances the asymptotic behavior $\chi_1(r)\approx
2\sin(r+\delta_1)/r$ gives
\begin{equation}\label{integral}
\int_0^\infty\frac{\chi_1(z)zdz}{z-\sqrt{\varepsilon}r}=2\pi\cos(\sqrt{\varepsilon}r+\delta_1)+O\left(\frac{1}{\sqrt{\varepsilon}r}\right).
\end{equation}
Obviously, this describes the molecule-atom channel
responsible for 3-body recombination. 
An outgoing wave of the
form $\exp(i\sqrt{\varepsilon}r)/r$ describing the dimer
and
atom flying apart, is obtained by adding a general solution
of the homogeneous form of Eq.(\ref{mainfinal}), which is
proportional to $f_{mol}(r)$, to
$f(r)$.
%The same result can be achieved if we
%take the integral (\ref{integral}) passing below the pole.

Eqs. (\ref{answer}), (\ref{integral}) and (\ref{Psi})
provide the
3-body wavefunction $\Psi(X)$ at distances
$1/\sqrt{\varepsilon}\ll\left| X\right|\ll 1/\sqrt{E}$,
where the inelastic dimer-atom channel is well separated
from the elastic one, and give the amplitude of
recombination. Then for the number of recombination
events
($\uparrow+\uparrow+\downarrow\,\longrightarrow\,\uparrow+\uparrow\downarrow$)
per unit time and unit volume in a gas we obtain
\begin{equation}\label{frequency}
\Omega_{\uparrow\uparrow\downarrow}=\alpha_\uparrow
(\overline{\epsilon_\uparrow}/\varepsilon)n_\downarrow
n_\uparrow^2,
\end{equation}
where $n_\downarrow$ and $n_\uparrow$ are the densities of
the fermionic components. The rate constant $\alpha_\uparrow
(\overline{\epsilon_\uparrow}/\varepsilon)$ is proportional
to the average kinetic energy of $\uparrow$-particles
$\overline{\epsilon_\uparrow}$, which
equals $3T/2$ in a nondegenerate gas, and $3T_F/5$ in a
deeply degenerate gas. The
coefficient $\alpha_\uparrow$ is given by
\begin{equation}\label{alpha}
\alpha_\uparrow=\frac{8\pi^3}{3}\frac{\hbar^5}{m_\uparrow^3\varepsilon^2}\lambda^2(1)\lambda^2(0)\lambda(-1)\sin^2\theta\tan^3\theta.
\end{equation}
Fig.(\ref{figalpha}) shows the dependence of the
dimensionless quantity
$\alpha_\uparrow m_\uparrow^3\varepsilon^2/\hbar^5$ on
$m_\uparrow/m_\downarrow$. In the case of
$m_\uparrow=m_\downarrow=m$ the quantity
$\alpha_\uparrow\approx 148\hbar
a^4/m$.
\begin{figure}
\hspace{-0.2cm}
\epsfxsize=\hsize
\epsfbox{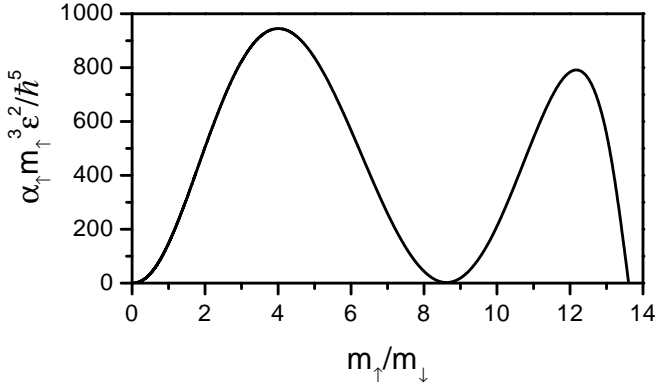}
\caption{\protect
The parameter $\alpha_\uparrow
m_\uparrow^3\varepsilon^2/\hbar^5$ versus $m_\uparrow/
m_\downarrow$.}
\label{figalpha}
\end{figure}The absence of 3-body recombination at the
points $\lambda(1)=0$ ($m_\uparrow/m_\downarrow=0$),
$\lambda(0)=0$ ($m_\uparrow/m_\downarrow\approx 8.62$) and
$\lambda(-1)=0$ ($m_\uparrow/m_\downarrow\approx 13.6$) is a
purely quantum phenomenon. From Eqs. (\ref{answer}) and
(\ref{integral}) we see that
for these mass ratios the free atom channel is decoupled
from the molecular recombination channel and the
interparticle interaction leads to elastic scattering only. 

The process
($\downarrow+\downarrow+\uparrow\,\longrightarrow\,\downarrow+\downarrow\uparrow$)
is described in a similar way. The results are given by
Eqs. (\ref{frequency}) and (\ref{alpha}), where one should
interchange the subscripts $\uparrow$ and $\downarrow$. In
a mixture of two
hyperfine components of the same
isotope, with equal densities $n_\uparrow=n_\downarrow=n/2$
and
$\overline{\epsilon_\uparrow}=\overline{\epsilon_\downarrow}=\overline\epsilon$,
the total loss rate of particles is
\begin{equation}\label{ndot}
-\dot n/n=Ln^2\approx 111\,(na^3)^2\,
\overline\epsilon/\hbar.
\end{equation}
This formula is valid provided the inequalities
$\overline\epsilon\ll\hbar^2/ma^2\ll\hbar^2/mR_e^2$ are
satisfied. In current experiments with
${}^{40}$K and ${}^6$Li one has $\overline\epsilon\sim
15\mu$K, 
$n\sim 10^{13}$cm$^{-3}$, and the scattering length can be
tuned 
using Feshbach resonances. For $a\approx 280$\AA $\,$
we calculate the rate constant $L\sim 10^{-25}$cm$^6/$s
and the loss rate (\ref{ndot}) is $\sim 10$ s$^{-1}$. 

In a trap with a barrier $\varepsilon_{\rm tr}$  higher than
the 
kinetic energy of dimers and fast atoms produced in the
recombination
process ($\sim\hbar^2/ma^2$), these particles remain trapped
and
Eq.(\ref{ndot}) overestimates the losses. Actually, the peak
loss
rate is reached at a finite scattering length 
$a\sim\sqrt{m\varepsilon_{\rm tr}}/\hbar$, not for
$a\rightarrow\infty$.
This is consistent with recent observations of particle
losses in trapped 
$^6$Li, where $a$ was tuned by a Feshbach resonance
\cite{Ketterle,Grimm,Thomas,Salomon}.

We emphasize the difference between
two-component Fermi gases and Bose gases with respect to
inelastic processes. Energy dependence of the recombination
in Fermi gases leads to decreasing $\overline\epsilon$
since particles with comparatively high kinetic energies
recombine. However, the degeneracy parameter decreases due
to a faster particle loss. Eq.(\ref{ndot}) gives an
estimate of the 3-body loss rate even if $a\sim R_e$ (i.e.
deep bound state). In Fermi gases the $a^6$ dependence
(\ref{ndot}) leads to a substantial suppression of the
recombination probability compared to the $a^4$ dependence
in Bose gases \cite{a4}. Further, in contrast to Bose
gases, cold
collisions between dimers and atoms have fewer inelastic
channels. Indeed, there are no weakly
bound Efimov $\uparrow\uparrow\downarrow$-states if
$m_\uparrow/m_\downarrow<13.6$, and the effective repulsion
between an atom and a dimer at low energies suppresses the
relaxation to deeper molecular states. In the case of a
deep trap and large $a$, where products of 3-body
recombination stay trapped, one expects a fast creation
of weakly bound molecules which can be further cooled. The
lifetime of the gas of these Bose dimers is sufficient to
observe, for example, a BEC.

We thank G.V.
Shlyapnikov, J.T.M. Walraven, M.A. Baranov, and M. Olshanii
for fruitful discussions. This work was
financially supported by the Stichting voor Fundamenteel
Onderzoek der Materie (FOM), by INTAS, and by the Russian
Foundation for Basic Studies.


\begin{references}
\bibitem{note} Resonant pair forces can also lead to a
3-body repulsion (see
\cite{EfimovMR}). This is the case for the $s$-wave
scattering of a $\uparrow$-fermion on a
$\uparrow\downarrow$-dimer.
\end{references}
\end{document}